\documentstyle[12pt]{article}
\begin{document}

\title{
$B \rightarrow X_{s} \tau^{+} \tau^{-}$ in the flipped SU(5) model}
\author{Chao-Shang Huang and Qi-Shu Yan\\
Institute of Theoretical Physics, Academia Sinica.\\
P.O. Box $2735$, Beijing $100080$, P.R. China}
\date{}
\maketitle
\vskip 2.5cm
\begin{minipage}{13 cm}
\begin{center} Abstract \end{center}
We perform a detailed study of the inclusive branching ratio and the forward-backward
asymmetry of the rare B decay process $B \longrightarrow X_{s} \tau^{+} \tau^{-}$
in the flipped SU(5) model, a predictive model which has only two free
parameters, plus the sign of the higgs mixing mass $\mu$, to describe mass spectrum and mixings of about $30$ sparticles. In contrast to
other works dealing with such a topic, our study has included the contributions of 
the neutral higgs bosons through penguin diagrams to this process under the context of supersymmetry.
In some regions of the parameter space, due to the substantial enhancement effects coming
from supersymmetry, these contributions become quite significant and could make the flavor change process
$B \longrightarrow X_{s} \tau^{+} \tau^{-}$ a microscope
to probe the regions of large tan$\beta$ and a window to gain an insight into new physics
beyond the standard model(SM).
\end{minipage}

\newpage

\section{Introduction}
The standard model(SM) is a successful effective low energy theory at energy
scales up to 100 GeV; its predictions
are in agreement with almost
all experimental tests. Nevertheless, there exist many theoretical problems
waiting for explanations. For example, where do Yukawa couplings
originate from? Why are there three, not other numbers of, generations?
Supersymmetry
is an excellent candidate which offers a scheme to embed
the SM in a more fundmental theory with which
these theoretical problems can be hopefully explained.
Supersymmetry has many good features. It ingeniously
tackles the abominable gauge hierarchy problem from which ordinary
grand unification theories suffer \cite{susy2}, provides a mechanism 
that supergravity models always share \cite{susy3} which
dynamically breaks the electroweak symmetry via radiative corrections and avoids
the arbitrary procedure to break $SU(2) \times U(1)$ group by hand, furnishs an elegant
framework to unify gravitational, strong and electroweak interactions \cite{susy4}, naturally proposes
the lightest supersymmetric particle(LSP) as a candidate of the dark matter in astrophysics and cosmology \cite{susy5},
and even can explain all parameters and properties of SM in
principles with only one input, as many superstring models have demonstrated \cite{rev2}. For all these compelling features,
supersymmetry faces its biggest problem: so far there are no decisive
evidences affirming its existence.

At present, in spite of lacking definite supports from experiments, supersymmetry
could be understood in an experimental perspective. Upon various possible new physics beyond SM,
latest experimental
results impose very severe constraints; supersymmetry
must confront these tests, as summarized in \cite{bll1}. 
About SUSY we know: 1)charged SUSY particles are heavier than 65 GeV, as given by the
experiments at LEP \cite{exp4}; 2)the gluino mass is larger than 100 GeV, a bound given
by Fermilab Tevatron collider \cite{exp3}; 3)all sneutrino masses are no less than 41 GeV \cite{exp1};
4)there should exist no seeds to break charge and color symmetry \cite{exp5}; 5)the lightest
supersymmetric particle is neutral; 6)the width of the process $Z \rightarrow \chi
\chi$ is less than 8.4 MeV, and branching ratios of $Z \rightarrow \chi
\chi^{\prime}$ and $Z \rightarrow \chi^{\prime} \chi^{\prime}$ are less than 
$2 \times 10^{-5}$; 7)the sparticles masses should not be much larger than 1TeV \cite{susy1},
otherwise the higgs bosons loop corrections would be too large and gauge hierarchy
problem arise again;
8)the most stringent constraint is brought in by the branching ratio of the rare decay
$b \rightarrow s \gamma$, the value is determined to be $(2.32 \pm 0.57 \pm
0.35 \times 10^{-4})$ with the upper bound $4.2 \times 10^{-4}$ and lower bound
$1 \times 10^{-4}$ at CLEO \cite{exp2}.

There are many supersymmetric models to be tested.
For experimentalists, to test a generic supersymmetric model, the minimal supersymmetrically
extended standard model without constraints on soft breaking parameters for instance, makes no much sense,
because so many possible explanations for a test can be offered by the model 
that no significant
predictions can be made for other tests. In contrast to this kind of models, a well
theoretically motivated model is highly predictive, therefore easier to be proven true or false,
and so is favored by experimentalists. Among various SUSY models, the flipped SU(5) model \cite{fsu55, bsg9},
one of a class of string-inspired supergravity models, is such a predictive
model. In the model the proton decay and cosmological constraints are satisfied automatically
and there are only two free parameters
(plus the sign of the higgs mixing mass $\mu$): 
the gaugino mass evaluated at the unification scale $m_{1/2}$ and the ratio of the
higgs vacuum expectation values tan$\beta$. Constraints from phenomenology listed
above greatly cut down the parameter spaces permitted by the model. In this note
we shall study $B \rightarrow X_s \tau^{+} \tau^{-}$ in the model.

As
analyzed in literatures \cite{bsg1,bsg2}, $b \rightarrow s \gamma$ puts a stringent
constraint on various possible new physics beyond SM.
For example, in the two-higgs-doublet model, it requires a larger charged higgs mass.
In SUSY, this restrictive condition can be relaxed or be grimmer
due to the either destructive or constructive interferences between the contributions
of charginos and those of $H^{\pm}$ and $W^{\pm}$ \cite{bsg9,bsg1,bsg2}.
In the flipped SU(5) model, for the case $\mu\!>\!0$, it can drastically reduce the parameter
space, especially the regions of permitted tan$\beta$, 
while for the case $\mu\!<\!0$, it imposes
no such a stringent constraint upon tan$\beta$ \cite{bsg9}.

Compared with the process $b \rightarrow s \gamma$ whose magnitude of branching ratio(BR) is 
$10^{-4}$, observing another rare B decay process $b \rightarrow s 
l^{+} l^{-}$ (its BR is $10^{-6}$\cite{bsg3})
is not available at present. But with the advent of the construction of B factories,
this process
is also quite promising and, because of its sensitivity to new physics, has been proposed as one
powerful piece of the arsenal to
discern supersymmetry from SM \cite{bsg3,bll2}. It was found in the minimal supergravity model that this process is strongly
correlated with $b \rightarrow s \gamma$ 
and in some regions of the parameter space the branching ratio of it can be
enhanced by about $50\%$ compared with SM \cite{bll1}. All the considerations above have
ignored the contributions from exchanging neutral higgs bosons(NHB). In SM, they can
be safely neglected owing to the smallness of $m_{l}/m_{w}$ (l=e, $\mu$, $\tau$).
But it is unjustifiable to omit them when beyond SM and
for the process
$b \rightarrow s \tau^{+} \tau^{-}$. As pointed out in \cite{bll3}, owing to 
the enhancement effects of the large tan$\beta$
in two-higgs-doublet model,
they become quite sizable and thus greatly influence the invariant mass distribution 
and forward-backward asymmetry of $b \rightarrow s \tau^{+} \tau^{-}$. Under the context of SUSY,
as we shall show below, a more significant enhancement effect of large tan$\beta$ 
coming from chargino-stop loops can even make them
dominant contributions so that in some regions of the parameter space
the branching ratio
of $b \rightarrow s \tau^{+} \tau^{-}$ 
can be enhanced by 200\% in both scenarios when compared
with SM. Therefore, it is possible that the first distinct signs of supersymmetry
could come from deviations in $b \rightarrow s l^{+} l^{-}$,
especially, for l=$\tau$.

\section{A brief description of the model}
The detailed description of the string inspired flipped SU(5) model
can be found in \cite{fsu55}. Here we only discuss two aspects of it
which are relevant to the study of $B \rightarrow X_{s} \tau^{+} \tau^{-}$ in this paper.
One is about the unification
of gauge couplings at string determined scale, the other the number of the
free parameters.

It is well known that the scale of gauge coupling unification derived from string theory is about $7 \times 10^{17}$ GeV \cite{fsu56,fsu57},
much larger
than the experimentally determined value $2 \times 10^{16}$ GeV, when assuming the particle
content of the minimal
supersymmetric standard model. Many approaches have been proposed to
reconcile the discrepancy between these two scales \cite{rev4}. In the flipped $SU(5)$
grand unified theory, this problem was solved by introducing one pairs of gap
particles(GPs) which can be economically embedded into the representations of $SU(5)
\times U(1)$\cite{fsu55}. They decouple from interactions later than heavier particles and bridge the 
gap between these two scales. 

Universal soft supersymmetric terms can be derived from local and global
supersymmetry breaking. Four free parameters, all are evaluated at unified scale,
are introduced here: $m_{1/2}$, the mass
of gauginos; $A$, the trilinear couplings; $B$, the 
bilinear couplings; $m_{0}$, the universal masses for all scalars.
String-inspired relations lead to two supersymmetry breaking scenarios:
1)the moduli scenario which requires $m_{0}$=A=0 \cite{susy3}; 2)the dilaton scenario which
requires $m_{0}$=$m_{1/2}/{\sqrt{3}}$, A=$-m_{1/2}$ \cite{fsu54}. Thus only B and $m_{1/2}$
are the free soft breaking parameters which will be further constrainted.

Electroweak symmetry is dynamically broken at low energy
due to the fact that the mass of one of the two higgs fields becomes negative
because of the effect of the large value of the top's Yukawa coupling. This is known as the radiative electroweak symmetry
breaking mechanism.
Minimal conditions of the scalar potential give
\begin{eqnarray}
& & \sin 2\beta=\frac{-2B\mu}{{m_{H_1}}^2+{m_{H_2}}^2+2 \mu^2} \nonumber \\
& & {\rm tg}^2 \beta=\frac{{m_{H_1}}^2+\mu^2+m_z^2/2}{{m_{H_2}}^2+\mu^2+m_{z}^2/2}
\end{eqnarray}
which go a further step to reduce the number of free parameters, for $B$ and $\mu^{2}$
can be solved out from these two equations while leaving the sign of the higgs
mixing parameter $\mu$ undetermined. Thus only two free parameters ($m_{1/2}$ and
tan$\beta$) survive
to completely describe mass spectra and mixings of about 30 sparticles.

We use eq.(1) and the renormalization group equations(RGEs)
of masses and couplings \cite{rge} to calculate mass spectra and mixings.
In order to simplify the computational procedure, we neglect
the Yukawa couplings of the first two
generations and the effects of GPs. A detailed analysis which includes these effects will be published elsewhere.
With the constraints from phenomenology we take the range of tan$\beta$ as from 2 to 40, and $m_{1/2}$
from 100 GeV to 400 GeV. The constraint from $b \rightarrow
s\gamma$ will be taken into account later on in our analysis.

\section{The formula for $B \rightarrow X_{s} \tau^{+} \tau^{-}$}
Inclusive decay rates of heavy hadrons can be calculated in heavy quark effective
theory and it has been shown that the leading terms in $1/m_Q$ expansions
turns to be the decay of a free (heavy) quark and corrections stem from
the order $1/m_Q^2$ \cite{hqet}. In what follows we shall calculate the leading
term. 

Under the context of supersymmetry, there are total five classes of loops
contributing to the flavor changing process $b \rightarrow s \tau^{+} \tau^{-}$
as well as to $b \rightarrow s \gamma$ by exchanging: W boson and u-type quarks,
charged higgs boson and u-type quarks, charginos and u-type squarks, neutralinos
and d-type squarks, gluinos and d-type squarks. Since the flavor mixings between
the third and the other two generations are small, contributions from the last
two cases are negligible so that we only consider contributions from 
the first three classes. The additional Feynman diagrams are shown
in fig.1. 

With Feynman rules at electroweak scale, as given in \cite{feyman},
effective weak Hamiltonian can be obtained
\begin{eqnarray}
H_{eff} &=& \frac{4G_{F}}{\sqrt{2}} V_{tb} V_{ts}^{*} (\sum_{i=1}^{10}
    C_{i}(\mu) O_{i}(\mu) + \sum_{i=1}^{10} C_{Q_i}(\mu) Q_i(\mu))
\end{eqnarray}
where $O_i$ and $Q_i$ are given in \cite{bll4} and \cite{bll3} respectively. The coefficients
$C_i(m_{W})$ can be found in \cite{bsg3}. We calculate
the coefficients $C_{Qi}(m_W)$ in SUSY models and the results are:
\begin{eqnarray}
C_{Q1}(m_W) &=& \frac{m_b m_{\tau}}{4 m_{h^0}^2 \sin^2\theta_W} {\rm tg}^2 \beta \{
 (\sin^2\alpha + h \cos^2\alpha) [ \frac{1}{x_{Wt}}(f_1(x_{Ht})-f_1(x_{Wt}))\nonumber\\
   &+&\sqrt{2} \sum_{i=1}^{2} \frac{m_{\chi_{i}}}{m_W} \frac{U_{i2}}{\cos \beta} (- V_{i1} f_1(x_{\chi_{i}{\tilde q}})
    +\sum_{k=1}^{2}\Lambda(i,k)T_{k1}f_1(x_{\chi_{i} {\tilde t_k}}))\nonumber\\
   &+&(1+\frac{m_{H_{\pm}}^2}{m_W^2}) f_2(x_{Ht},x_{Wt})]   
    -\frac{m_{h_0}^2}{m_W^2} f_2(x_{Ht},x_{Wt})\nonumber\\
   &+& 2 \sum_{ii'=1}^2 (B_{1}(i,i') \Gamma_{1}(i,i')+A_{1}(i,i') \Gamma_{2}(i,i'))\}\nonumber\\
C_{Q2}(m_W) &=&- \frac{m_b m_{\tau}}{4 m_{A^0}^2 \sin^2\theta_W}{\rm tg}^2 \beta\{\frac{1}{x_{Wt}}(f_1(x_{Ht})-f_1(x_{Wt}))
    +2 f_2(x_{Ht},x_{Wt})\nonumber\\
    &+& \sqrt{2} \sum_{i=1}^{2} \frac{m_{\chi_{i}}}{m_W} \frac{U_{i2}}{\cos \beta} (- V_{i1} f_1(x_{\chi_{i} {\tilde q}})
    +\sum_{k=1}^{2}\Lambda(i,k)T_{k1}f_1(x_{\chi_{i}{\tilde t_k}})) \nonumber\\
    &+& 2 \sum_{ii'=1}^2 (-U_{i'2}V_{i1} \Gamma_{1}(i,i')+U^{*}_{i2}V^{*}_{i'1} \Gamma_{2}(i,i'))\}
\end{eqnarray}
where
\begin{eqnarray}
B_{1}(i,i')&=&(-\frac{1}{2} U_{i'1}V_{i2}\sin2\alpha (1-h)+U_{i'2}V_{i1} (\sin^2\alpha + h \cos^2\alpha))\nonumber\\
A_{1}(i,i')&=&(-\frac{1}{2} U_{i1}^{*}V_{i'2}^{*}\sin2\alpha (1-h)+ U^{*}_{i2}V^{*}_{i'1}(\sin^2\alpha + h \cos^2\alpha))\nonumber\\
\Gamma_{1}(i,i')&=& m_{\chi_{i}} m_{\chi_{i'}} U_{i2} (- \frac{1}{{\tilde m}^2}f_2(x_{\chi_{i}{\tilde q}},x_{\chi_{i'}{\tilde q}}) V_{i'1} +\sum_{k=1}^2 \frac{1}{m_{\tilde t_{k}}^2}\Lambda(i',k) T_{k1} f_2(x_{\chi_{i}{\tilde t_k}},x_{\chi_{i'}{\tilde t_k}}))\nonumber\\
\Gamma_{2}(i,i')&=&U_{i2} (- f_2(x_{\chi_i{\tilde q}},x_{\chi {i'}{\tilde q}}) V_{i'1} +\sum_{k=1}^2 \Lambda(i',k) T_{k1} f_2(x_{\chi_i {\tilde t_k}},x_{\chi_{i'}{\tilde t_k}}))\nonumber\\
\Lambda(i,k)&=&V_{i1}T_{k1} - V_{i2}T_{k2}\frac{m_t}{\sqrt{2} m_W \sin\beta}\nonumber\\
f_1(x_{ij})&=&1 -\frac{x_{ij}}{x_{ij}-1}ln ~ x_{ij}+ln ~ x_{Wj}\nonumber\\
f_2(x,y)&=&\frac{1}{x-y}(\frac{x}{x-1} ln ~ x-\frac{y}{y-1} ln ~ y)\nonumber\\
x_{ij}&=&m_i^2/m_j^2 
\end{eqnarray}
with $m_i$ being the mass of the particle i,
and
\begin{eqnarray}
C_{Q3}(m_w)&=&\frac{m_b e^2}{m_{\tau} g^2}\{C_{Q1}(m_w)+C_{Q2}(m_w)\}\nonumber\\
C_{Q4}(m_w)&=&\frac{m_b e^2}{m_{\tau} g^2}\{C_{Q1}(m_w)-C_{Q2}(m_w)\}\nonumber\\
C_{Qi}(m_w)&=&0, (i=5,\cdots 10)
\end{eqnarray}

In eqs.(3) and (4), $U$ and $V$ are matrices which diagonalize the mass matrix of charginos,
$T$ is the matrix reflecting the mixing of stops ${t_R}$ and ${t_L}$, $m_{\chi_j}$
denote the chargino masses, $\tilde{m}$ is the average mass of u-type squarks ${\tilde q}$ of the
first two generations, h is the square of the ratio of the mass of
$h^0$ to the mass of $H^0$ and $\alpha$ is the mixing angle of neutral
components of the two higgs doublets in the model. And in eq.(3) we have 
omited less important terms because they are numerically negligible compared
to those given in eq.(3) when tan$\beta \geq 20$.\\
Considering QCD corrections and evolving these coefficients down
to the scale we interest in, the effective Hamiltonian results in the
following matrix elements for $B \rightarrow X_{s} \tau^{+} \tau^{-}$
\begin{eqnarray}
M &=& \frac{G_F\alpha}{\sqrt{2} \pi} V_{tb} V_{ts}^{*} 
      [C_8^{eff} {\bar s_L}\gamma_{\mu}b_L{\bar \tau}\gamma^{\mu}\tau + 
    C_9 {\bar S_L}\gamma_{\mu}b_L{\bar \tau}\gamma^{\mu}\gamma^5\tau \nonumber \\ 
    & & +2C_7m_b{\bar s_L}i\sigma^{\mu\nu}\frac{q^{\mu}}{q^2}b_R{\bar \tau}
    \gamma^{\nu}\tau + C_{Q_1}{\bar s_L}b_R{\bar \tau}\tau + 
    C_{Q_2}{\bar s_L}b_R{\bar \tau}\gamma^5\tau ]
\end{eqnarray}
here these coefficients are evaluated at $\mu$=$m_b$. $C_8^{eff}$ is given as\cite{bll4}:
\begin{eqnarray}
C_8^{eff} &=& C_8 +\{g(\frac{m_c}{m_b},s)+\frac{3}{\alpha^2}k\sum_{V_{i}}\frac{\pi M_{V_{i}} \Gamma(V_{i} \rightarrow \tau^{+} \tau^{-})}{M_{V_{i}}^2 -q^2-iM_{V_{i}}\Gamma_{V_{i}}}\}(3C_1 +C_2)
\end{eqnarray}

From eq.(6), by integrating the angle variable of the double differential 
distributions from 0 to $\pi$, the invariant dilepton mass distributions
can be calculated and given below
\begin{eqnarray}
\frac{{\rm d}\Gamma(B\rightarrow X_s\tau^{+}\tau^{-})}{{\rm d}s}
 &=& B(B\rightarrow X_c l {\bar \nu}) \frac{{\alpha}^2}
 {4 \pi^2 f(m_c/m_b)} (1-s)^2(1-\frac{4t^2}{s})^{1/2}
 \frac{|V_{tb}V_{ts}^{*}|^2}{|V_{cb}|^2} D(s) \nonumber \\
 D(s) &=& |C_8^{eff}|^2(1+\frac{2t^2}{s})(1+2s)
      + 4|C_7|^2(1+ \frac{2t^2}{s})(1+\frac{2}{s}) \nonumber \\
    & &  + |C_9|^2 [ ( 1 + 2s) + \frac{2t^2}{s}(1-4s)]
      +12 {\rm Re}(C_7 C_{8}^{eff*})(1+\frac{2t^2}{s}) \nonumber \\
  & & + \frac{3}{2}|C_{Q_1}|^2 (s-4t^2) + \frac{3}{2}|C_{Q_2}|^2s
      + 6{\rm Re}(C_9 C_{Q_2}^{*}) t
\end{eqnarray}
where s=$q^2/m_b^2$, t=$m_{\tau}/m_{b}$, $B(B\rightarrow X_c l {\bar \nu})$ is the branching ratio
which takes as 0.11, $f$ is the phase-space factor and f(x)=$1-8 x^2+8 x^6
-x^8-24 x^4 \ln ~ x$.
The forward-backward asymmetry of the lepton in the process has also been given
\begin{eqnarray}
A(s) &=&- 3 (\frac{1-4t^2}{s})^{1/2}E(s)/D(s)\nonumber\\
E(s)&=& {\rm Re}(C_8^{eff}C_9^{*})s 
  + 2{\rm Re}(C_7C_9^{*})
 + {\rm Re}(C_8^{eff}C_{Q_1}^{*})t
  + 2{\rm Re}(C_7C_{Q_1}^{*})t
\end{eqnarray}
In eqs.(8) and (9) the mass of strange quark has been neglected.

\section{Numerical analysis}
The constraints on the $SU(5)\times U(1)$ model from $b\rightarrow s \gamma$ have been
in detail studied in the ref.\cite{bsg9}. It is shown that there is an
upper bound on tan$\beta$: tan$\beta \leq 25$ for $\mu\!>\!0$ and 
there is no analogous bound for $\mu\!<\!0$. The reason is that
when $\mu\!>\!0$, the supersymmetric
contributions of chargino loops have the same sign with contributions 
of w and charged higgs
loops and interfere constructively, therefore tan$\beta$ is greatly 
constrained and the
large values of it are not favored; when $\mu\!<\!0$, contributions of SUSY have
a opposite sign and interfere destructively, therefore the constraints are relaxed.
Because we are interested in the region of large tan$\beta$,
we only consider the case of $\mu\!<\!0$ in the letter. 

We analyze the effects of the supersymmetric flipped SU(5) model 
to the experimental observables --- the invariant
mass distribution and backward-forward asymmetry. It is obvious from eq.(3) that
$C_{Q_{i}}$ (i=1, 2) can reach a value of order one only when tan$\beta$ is large enough
(say, $\geq 20$) due to the smallness of $m_b m_{\tau}/m_h^2$ ($h=h^0$, $A^0$).
For larger values of tan$\beta$, the contributions of exchanging NHBs dominate,
which is shown
in fig.2. From the fig.2 one can see
the deviations from SM are quite substantial when tan$\beta$=25(30) in moduli(dilaton) scenario.
The enhancement of the invariant mass distributions is about 200\% in both scenarios.
The backward-forward asymmetry is greatly modified 
in both scenarios.
The predictions without including the
contributions of exchanging NHBs are also shown in fig.2
in order to compare. It is obvious from fig.2 that
the invariant mass distribution without including the contributions of
exchanging NHBs is almost equal to
that in SM because for the values of tan$\beta$ and $m_{1/2}$ indicated in the fig.2 
contributions of SUSY without including those of exchanging NHBs almost do not change
the values of $C_i$(i=7, 8, 9)
in SM, while for some other values of tan$\beta$ and $m_{1/2}$,
the contributions can change the sign of $C_7$ and consequently enhance the invariant mass distribution by about 50\%
compared to SM, similar to the conclusion made in the reference\cite{bll1}.

\section{Conclusion}
In this paper, we have studied the rare B decay process 
$B \rightarrow X_s \tau^{+} \tau^{-}$ in the flipped SU(5)
model. In particular, the contributions of exchanging
neutral higgs bosons
is intensively analyzed. It is found that 
in the regions where $m_{1/2}$ is moderate and tan$\beta$ is large, the branching
ratio of $B \rightarrow X_s \tau^{+} \tau^{-}$ is enhanced by about 200\%
compared to the SM and the forward-backward
asymmetry is also significantly different from the SM.
Therefore, the process $B \rightarrow X_s \tau^{+} \tau^{-}$
is of a good probe to investigate the model and search for new physics.

\vskip 1cm
\begin{center}
{\large \bf Acknowledgement}
\end{center}
One of the authors (Q.S. Yan) would like to thank W. Liao, H.G. Yan and G.X. Ju
for their generous helps and useful discussions. This work was supported in part
by the National Natural Science Foundation of China and partly supported by
Center of Chinese Advanced Science and Technology(CCAST).

\vskip 6cm
\begin{center}
{\Large \bf Figure Captions}
\end{center}
\vskip 0.5cm
Fig.1 The additional Feynman diagrams corresponding to exchanging NHBs in
SUSY model, where wavy lines represent the charged bosons: $W^{\pm}$,
$H^{\pm}$, $\tilde{t_i}$, internal solid lines t quark or charginos, and
dashed lines NHBs $H^0$, $h^0$, $A^0$.
\vskip 0.5cm
\noindent Fig.2 $d\Gamma/ds$ and A(s) for the case $\mu\!<\!0$, a) tan$\beta$=25 and
$m_{1/2}$=300 GeV in the moduli scenario and b) tan$\beta$=30 and 
$m_{1/2}$=350 GeV in the dilaton scenario.
The solid , dashed and dotted lines represent the predictions of the
flipped SU(5), the flipped SU(5) without contributions of NHBs and
SM respectively.
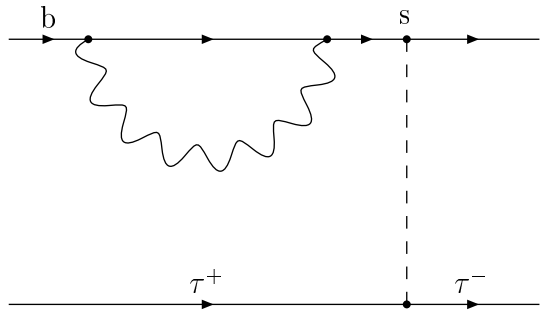
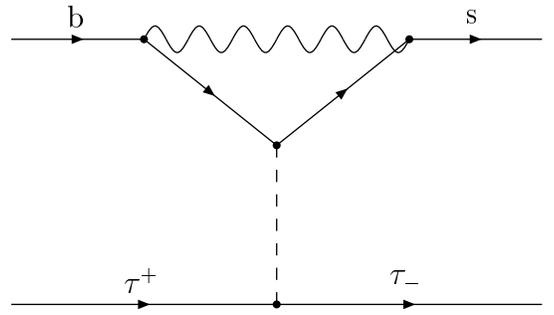
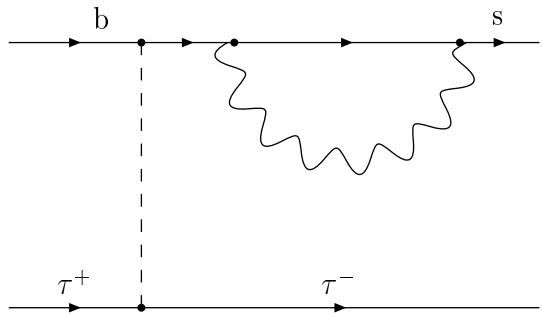
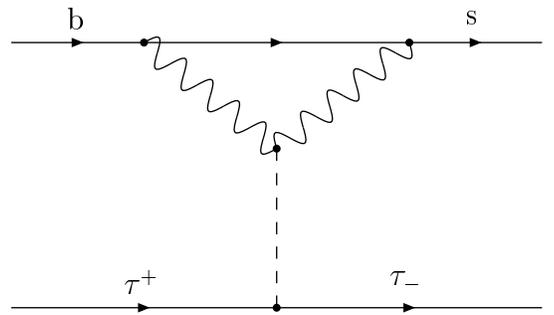
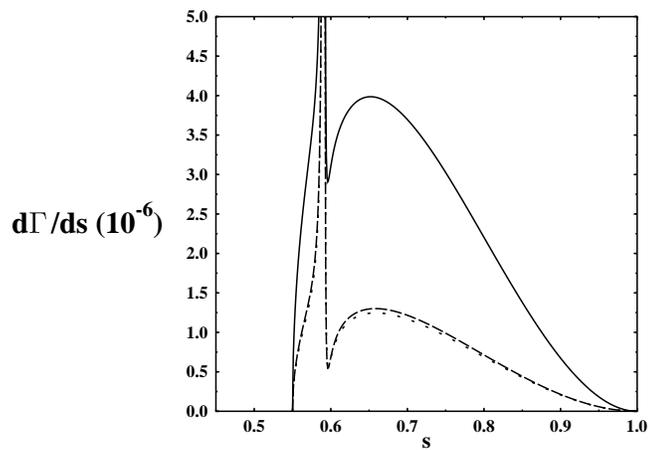
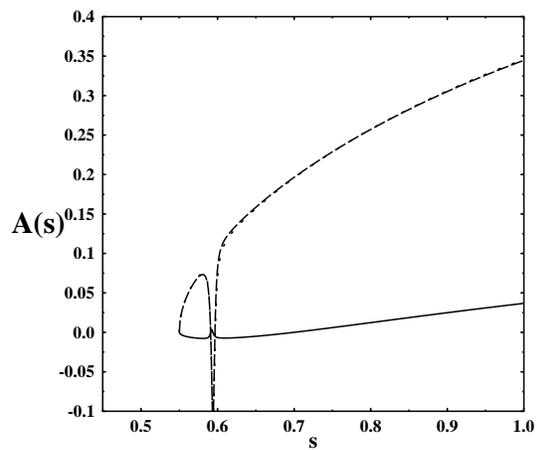
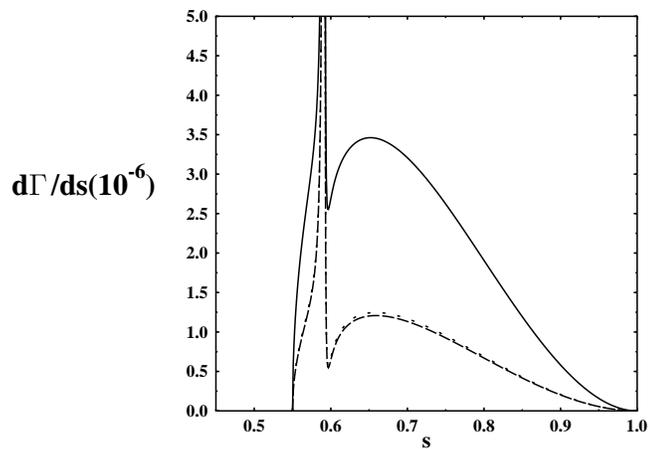
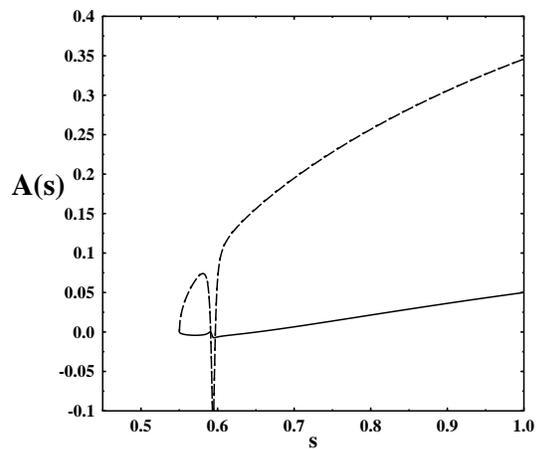
\end{document}